\documentclass[10pt,conference]{IEEEtran}
\IEEEoverridecommandlockouts

\usepackage{graphicx}
\usepackage{amsmath}
\usepackage{amssymb}
\usepackage{url}
\usepackage[hidelinks]{hyperref}

\begin{document}

\title{``What's That Sound?'': A Versatile, Robust, and Lightweight Convolutional Transformer for Environment Sound Recognition\thanks{Originally presented at the 2022 International Young Researchers' Conference.}}

\author{\IEEEauthorblockN{Julia Huang}
\IEEEauthorblockA{Northville High School, Northville, USA\\
Email: juliah6169@gmail.com}}

\maketitle

\begin{abstract}
The conventional hearing aid is both costly and limited in usage, as it is not intended to detect non-speech audio. Our objective is to develop a machine learning solution to provide a more accurate and affordable mechanism to identify surrounding sounds to improve the safety of the hearing impaired, i.e., if a car is honking behind pedestrians, or a gunshot is fired, and they need to move away from the source. By adding randomized augmentations to audio, concatenating a Mel-Frequency Cepstral Coefficients (MFCCs) diagram and a log-mel Spectrogram, and including Convolutional Neural Networks (CNNs) in a Transformer architecture, the Randomized Audiomentational Layered Convolutional Transformers (RALCT) model efficiently extracts features from diversified audio representations. In addition, RALCT is small enough, with only approximately 310,000 parameters, to be deployed into mobile devices. Experimental results on the UrbanSound8K dataset resulted in an accuracy consistently over 93\% for all variations of RALCT with the highest at 94.56\%, reaching state-of-the-art levels. To leverage the capabilities of this technology, a mobile app is developed to be integrated with the model to provide real-time safety control. RALCT thus represents a robust, lightweight, affordable, and versatile deep learning tool to aid the navigation and safety of the hearing impaired.
\end{abstract}


\section*{Background}

Environmental sound classification (ESC) is a crucial task for aiding the hearing-impaired community. According to the World Health Organization, over 430 million people in the world are hearing impaired, and the current solution, the hearing aid, is both costly and limited in usage (``World Health Organization,'' n.d.). Costing between \$6,000 and \$8,000, hearing aids are virtually inaccessible to the low-income hearing-impaired community (Olson, 2021). In addition, they are usually designed to magnify human voices, not detect background noises; in fact, they amplify and reduce noise and human speech simultaneously, making it difficult to hear each sound type separately (Audiologist, 2020). Further, because a hearing aid is an electronic system based on acoustics, its performance will inevitably deteriorate over time (Fourie, 2018). Developing an AI-based framework lightweight enough to be deployed into a portable application to identify outdoor sounds, therefore, would be a more reliable and affordable solution for the hearing impaired.

\section*{Previous Work}

Traditional ESC machine learning solutions consist of straightforward audio deformation techniques, or augmentations, and a singular model. The most popular algorithm utilized in ESC is Convolutional Neural Networks (CNNs) due to its ability to learn localized features of input data indicative of certain categories by using a filter that scans features and generates a feature map, making these networks robust for classification tasks (Bonner, 2019).

\subsection*{CNNs}

Both audio augmentations and the CNN were used by Salamon \& Bello (2016) who applied them directly to raw audio files before converting them to a mel-spectrogram, a 2D representation of audio that records frequencies of an audio signal over time (Salamon \& Bello, 2016). The spectrogram is the first input of their CNN, resulting in a classification accuracy of 79\% on a multi-class audio dataset, UrbanSound8K (Gorgolewski, 2020). (To establish accurate comparisons with our method, all previous work mentioned in this paper utilize the same dataset.) However, this augmentation technique results in a limited scope of representations for audio files, as changes are only applied once in a fixed way, which makes it difficult for their model to generalize patterns specific to each audio category to result in a high classification accuracy.

On the other hand, Palanisamy et al. (2020) presented an ensembled CNN architecture, which consists of merged individual models, with an 87.42\% accuracy (Palanisamy et al., 2020). They also used the transfer learning technique by pretraining their ensembled architecture on a larger image dataset, ImageNet. This way, they can set pretrained initial weights on their ensemble instead of random weights for audio classification. This approach often provides architectures with more information and helps models more robustly understand patterns in data from smaller datasets, hence resulting in higher accuracy (Gupta, 2017).

However, even with pretraining, the ensembled model still classified 12.58\% of audio files incorrectly, indicating that there may be a flaw within the CNN architecture itself. The CNN architecture does not record feature positions, reducing its understanding on how features are connected for input data that has time or location dependency, such as audio (Giacaglia, 2019).

\subsection*{Transformers}

In recent years, the Transformer architecture has shown great promise in speech and text applications. Its applications in audio are relatively new; nonetheless, it has achieved high accuracy. Guzhov et al. (2021) achieved 90.07\% on UrbanSound8K with their proposed Transformer, AudioCLIP (Guzhov et al., 2021). Ahmadian et al. (2021) also utilized a Transformer, the MhaNN, which received 92.2\% (Ahmadian et al., 2021). The Transformer enhances important parts of input, records feature locations, and establishes dependencies between parts of data to capture the entire structure of data.

However, purely using Transformers presents an issue at hand: if a spectrogram or a Mel-Frequency Cepstral Coefficients (MFCCs) representation, which converts audio into the frequency domain, has many time steps, then the mathematical operations of the Transformer's multihead attention layer becomes very computationally expensive ($O(n^2 \cdot d + n \cdot d^2)$) (Vaswani et al., 2017). As many Open-Source Machine Learning environments have a limited amount of Random Access Memory (RAM), it is not possible to train a Transformer on audio representations with a large time dimension size. Further, many Transformer models are inherently far too large. Though larger models can be powerful enough to learn detailed patterns, their bulky sizes consume significant amounts of storage; therefore, they are unable to be deployed into small IoT devices. For example, though AudioCLIP achieved 90.07\% accuracy, it totals to 30 million parameters, far too large to fit in on-chip storage.

\section*{Methodology}

\subsection*{Motivation of Approach}

Our research aims to overcome the above shortcomings and present a more practical, affordable, and robust machine learning algorithm that has the following benefits: it requires low cost and resources to set up and train, it is consistently accurate regardless of slight parameter changes, and it is lightweight enough to run real-time on mobile devices to detect outdoor audio in real time. We extend the third point by developing a mobile app prototype to integrate with our proposed algorithm. This combined solution could help hearing impaired users navigate busy roads and streets to avoid vehicle- and gun-related injuries, and it could also improve navigational safety in the long-term by allowing users to share incident locations and keep records of dangerous sounds on file.

Our major motivation for proposing an improved audio classification solution stems from possible, wide-ranging applications. In addition to helping the hearing impaired navigate busy outdoor streets, our solution can easily be extended to navigation inside buildings and in the wilderness with just training our model on another dataset, such as one containing indoor sounds like doorbell or nature sounds like snake hissing. Carrying just a phone containing our free audio classification app can potentially serve as a more convenient and accurate solution for up to 430 million hearing impaired users. In sum, our objective is to offer a scalable solution for a variety of outdoor scenarios.

\subsection*{Proposed Process and Model Architecture}

Our study includes processing a dataset, building and training a novel lightweight deep learning architecture, and developing a mobile app to integrate with the model.

To utilize both models' contrasting advantages, we combined the CNN and Transformer, taking advantage of the CNN's fast training time and robust ability to extract features while reducing computational complexity and the Transformer's ability to establish long-range dependencies between features (Dai et al., 2021). Because CNNs can take in a large number of time steps of a spectrogram or MFCCs representation and reduce their dimension sizes, the resulting representation can easily be transferred to the Transformer framework without sacrificing RAM or computational cost. Exploiting the strengths of both the CNN and Transformer enable our model to precisely identify and understand key features in audio and classify each sound with higher accuracy. We also utilized randomized audio augmentations and a combined MFCCs representation and log-mel spectrogram to retrieve even more diversified audio representations, helping our model more accurately interpret audio. Therefore, we propose the Randomized Audiomentational Layered Convolutional Transformers (RALCT). We believe our architecture is the first ever combined CNN and Transformer model for audio classification.

Our proposed RALCT model is built on the TensorFlow 2.8.2 framework and trained in the Google Colab Jupyter Notebook. Our only overhead purchase is Colab Pro Plus, which we utilize to access faster GPUs. The code of our process can be found in our GitHub repo at \url{https://github.com/TheClassicTechno/audioclassmodel}.

\begin{figure}[htbp]
\centering
\includegraphics[width=\linewidth]{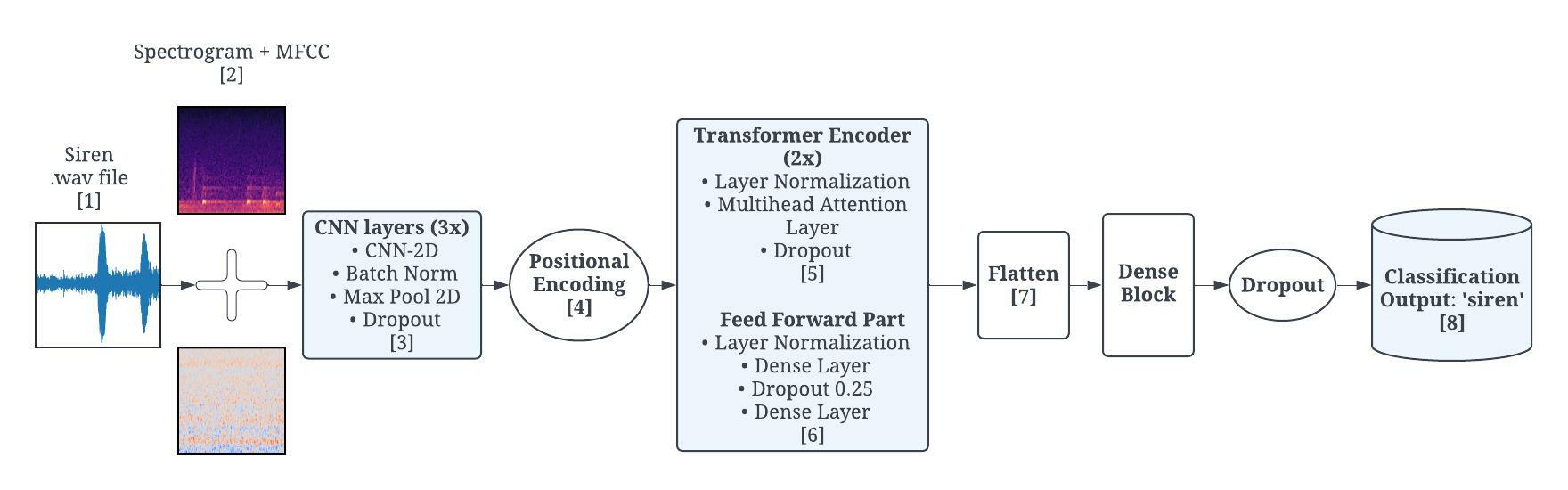}
\caption{The proposed RALCT architecture. The audio waveform is first resized to a length of 5 seconds then modified with randomized augmentations. In the first layer of RALCT, 1D audio is transformed into log-mel spectrogram and MFCCs representations, and CNN layers extract local features. Next, the representations are resized and passed into the Transformer for extracting features across time and frequency.}
\label{fig:architecture}
\end{figure}

In the first step, files from UrbanSound8K are modified with three randomized data augmentation methods before being processed. Utilizing randomized augmentations allows our model to see a large range of variations and representations for each audio file, ultimately improving pattern generalization and learning. Our technique randomly applies noise, pitch, and time shift from the Librosa library to training files during each epoch with an adjustable intensity for each augmentation. For time stretch, a random number is chosen to change the speed of a random sound file to between $0.8\times$ to $1.5\times$ of its speed. For pitch shift, a random function increases or decreases the pitch of a file by five semitones. Lastly, we add noise by creating a random array of values, multiplying the array by a small noise factor, and adding the array to the audio file. These augmentation techniques help avoid overfitting and allow models to be suitable to train on small datasets, such as UrbanSound8K. Since our process involves training our model numerous times with different adjustments of parameters to ensure robustness, each time the model is also exposed to a different augmented dataset. Lastly, the Randomized in RALCT is demonstrated as we provide customizable probability variables for each of the three augmentations. To elaborate, if we set the chances of pitch shift, time stretch, and noise addition to 0.2, each audio clip processed has a 20\% (0.2) chance to get augmented by one of the methods, 4\% ($0.2\times0.2$) chance to get modified by two, and so forth.

Next, we discuss the RALCT architecture from end to end, which is detailed in Figure~\ref{fig:architecture} and numbered by (\#) in the paper. First, RALCT takes in the augmented and processed sound files (1) as input, then computes them into a log-mel spectrogram and MFCCs diagram (2) that are separately normalized using Batch Normalization to help our model converge earlier and find optimal weights quicker. Afterwards, the combined representation is passed into the three CNN layers, each of which has a set number of filters that determine the number feature maps produced and the variety of ways to describe audio (3).

After each Convolutional (Conv.) Layer, a Dropout layer is added to prevent overfitting, or help the model learn features instead of memorizing. Dropout layers randomly set weights of select parts of audio to zero to reduce the model's dependency on specific neurons. This technique spreads the learning among all neurons in RALCT to allow every neuron to learn, making the architecture more robust. The Max Pooling layer reduces feature map dimensions by downsampling the combined MFCCs and spectrogram dimensions, reducing the number of parameters the model needs to learn and preventing computations from becoming too complex. Next, after the CNNs reduce the time dimension size of the concatenated 2D visual representation, it is added to a positional encoder (4) to help the model learn the location of each feature before being passed into two Transformer encoder blocks (5) (Kazemnejad, 2019).

The Transformer encoder maps audio into a higher dimensional space and records important features (Allard, 2019). Within the Transformer there is a multi-head attention layer, where each head attends to a different important part of the data. Dropout is used in between attention layers so RALCT can generalize patterns, and a residual connection is concatenated with the inputs to allow the model to decide the importance of previous features. The feedforward section (6) helps the model with memory retention, and its dense layers combine features with the most optimal audio representations chosen by the model. The Flatten layer (7) decreases dimensions of the feature map and squeezes the map into 1D numerical values to feed into the Dense layer, which contains hidden layers that try to fit the data and utilizes matrix multiplication in the Flatten layer. In the end, RALCT outputs one of the ten audio class names using a softmax activation function (8) (Nandi, 2021).

\section*{Experimental Procedure}

\subsection*{Dataset}

To evaluate RALCT's performance on outdoor sounds, we chose the largest and most comprehensive dataset publicly available to us, UrbanSound8K. This dataset consists of 8,732 files classified into 10 categories, including car horn, engine idling, gun shot, siren, etc. Each category contains 374 and 1,000 files with all tracks recorded in 22,050 or 44,100 Hertz (Hz) in mono 16-bit wav format. During preliminary training, we noticed a consistent low classification accuracy for certain classes, especially children playing, dog bark, and street music. Upon further inspection on the incorrectly classified sound clips, we discovered that approximately 11 of these clips were either mislabeled, contained multiple categories of sounds, or contained sounds that do not belong in any of UrbanSound8K's categories. UrbanSound8K contains only mono-labeled data (every file belongs to a single category), so we applied data cleaning to remove multi-label and no-label tracks and relabel files correctly. For each of the three augmentations, we set the chance variables to 0.2 (20\%) before processing 8,721 sound clips at a 44.1kHz sampling rate.

\subsection*{Training Setup}

80\% of tracks in the dataset are randomly selected for training and the remaining 20\% for validation. Our model's accuracy is computed by averaging the classification results of the 10 classes. We use the Adam optimizer to tune the model's parameters before training. In the training stage, we experiment with the initial learning rates $lr = 10^{-4}$ and $lr = 10^{-3}$. Instead of applying a constant learning rate for all 120 epochs, we utilize Cosine Decay on the $lr$, decreasing the starting rate incrementally. This technique is shown to aid optimization and generalization of patterns, as the initially large learning rate prevents models from learning noisy data and the decaying aids models in learning complex patterns in input data (You et al., 2019). To reduce chances of overfitting, we select batch sizes (bs) of 32 or 64 for each epoch.

At bs=32, our model takes 1.1385 hours to train with 34 seconds per epoch step, and at bs=64, RALCT takes 46.428 minutes to train with 23 seconds per epoch.

\section*{Results}

\begin{table}[htbp]
\caption{Accuracies of different architectures on UrbanSound8K compared with the most optimized RALCT version. All the models listed are either pure CNN or Transformer architectures besides MhaNN-SVM.}
\label{tab:comparison}
\centering
\includegraphics[width=\linewidth]{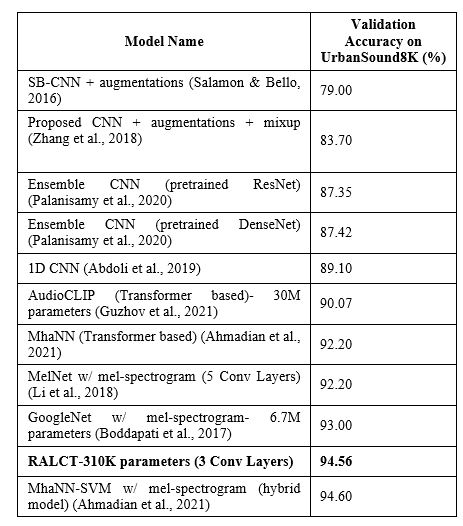}
\end{table}

\begin{table}[htbp]
\caption{Accuracies of different RALCT versions. Unless indicated in the table, both CNN and Transformer layers are included, the 3 Conv Layers have 32-64-32 filters each, Conv dropouts are 0.1, augmentation chances are 0.2, and both the head\_size and mlp\_units (multiperceptron units) of the Transformer are 64. Note: the first three trials are not official RALCT architectures due to not having randomized augmentations and/or not including both CNNs and Transformers.}
\label{tab:ablation}
\centering
\includegraphics[width=\linewidth]{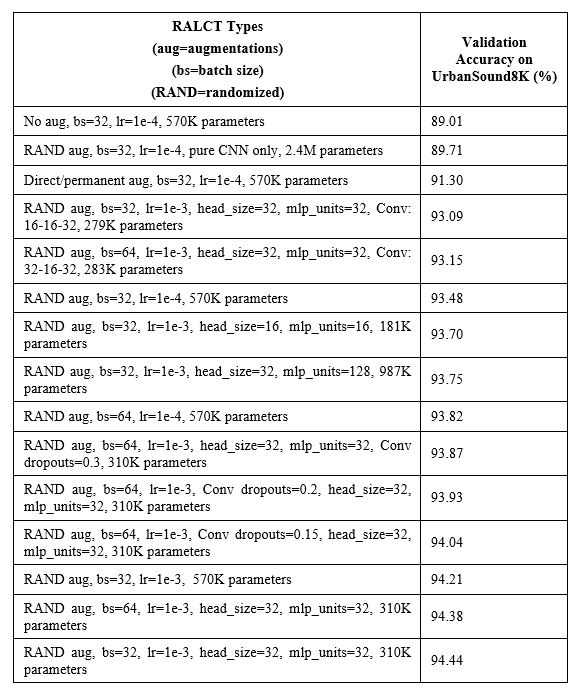}
\end{table}

Table~\ref{tab:comparison} displays the performance of previous works' CNNs and Transformers compared with our most optimized RALCT model in terms of overall percentage validation accuracy. RALCT outperforms all other models due to the disadvantages of using only CNNs or Transformers. Though MhaNN achieved similar accuracy to RALCT, it achieved so after 400 epochs, while RALCT only needed 119 epochs. Similarly, though GoogleNet and MelNet, CNN-only architectures, resulted in performances close to RALCT, they are significantly larger, making them much more computationally expensive and less able to be integrated within small devices, which is the primary objective of our research.

All previous work used pure CNNs or Transformers except the combined MhaNN-SVM, which only surpassed RALCT by 0.04\%. MhaNN-SVM uses a support vector machine (SVM) as their classifier instead of a CNN. SVM has demonstrated good performance in recognizing nonlinear and high-dimensional patterns, such as in audio, and so we suspect that this SVM addition may be the reason their model achieved comparable accuracy to RALCT, as their MhaNN only achieved 92.20\%. While MhaNN-SVM also reached 94.60\% after 400 epochs, RALCT reached comparable accuracy after only 119 epochs.

Table~\ref{tab:ablation} demonstrates accuracies of different forms of the RALCT algorithm and contains Ablation Studies. Even with different parameter modifications, validation accuracies are consistently above 93\%. We observe that RALCT with an intermediate number of parameters (310K) performed the best. In addition, we observe that increasing dropout in Conv layers also hurts accuracy; we suspect this is because dropping too many weights prevents RALCT from understanding critical connections between data. Lastly, we conclude that a larger initial learning rate ($10^{-3}$) improves performance because it causes the model to learn faster.

\begin{figure}[htbp]
\centering
\includegraphics[width=\linewidth]{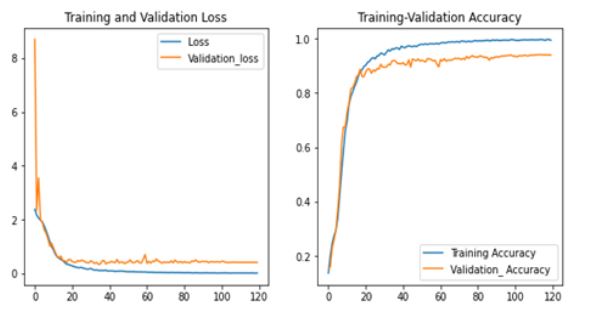}
\caption{Validation and training losses (left) and accuracies (right), respectively, compared side by side. The validation curve runs parallel to the training curve with little fluctuation. The left graph displays a 0.0160 training loss and 0.3700 validation loss; the right displays a 99.54\% training accuracy and 94.56\% validation accuracy.}
\label{fig:loss}
\end{figure}

\begin{figure}[htbp]
\centering
\includegraphics[width=\linewidth]{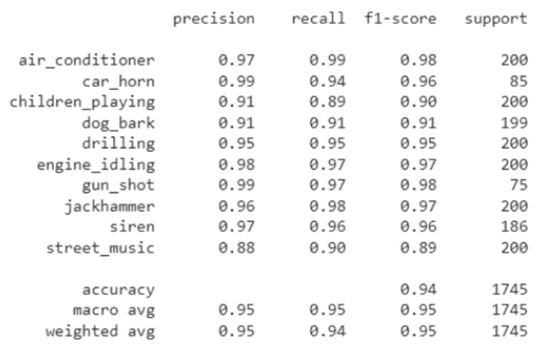}
\caption{Precision, Recall, and F1-Score table, displaying accuracies for each category.}
\label{fig:classreport}
\end{figure}

\begin{table}[htbp]
\caption{The confusion matrix for RALCT (most optimized version). The y-axis describes true classes while the x-axis contains predicted classes.}
\label{tab:confusion}
\centering
\includegraphics[width=\linewidth]{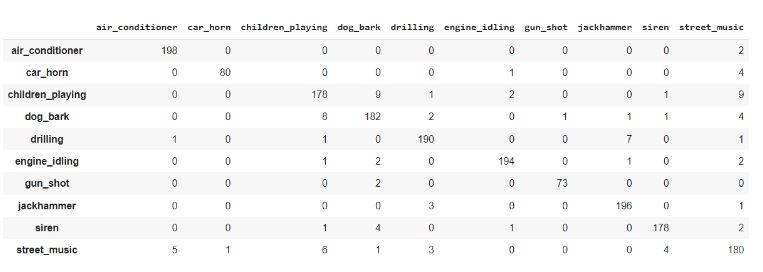}
\end{table}

Figure~\ref{fig:loss} presents training and validation results for overall loss and classification accuracy. Our most optimized RALCT specifically reached 94.56\% at epoch 119. The x-axis on the graphs displays the number of epochs completed and the y-axis displays loss on a scale from 0 to 4 and accuracy from 0.0 to 1.0 (100\%). In later stages, training and validation curves run parallel to each other, remaining consistent, proving the robustness of our model.

Figure~\ref{fig:classreport} provides three different metrics to provide a more reliable performance metric for each category. The RALCT model performed the best in the gunshot category because its quick sound is unique and distinctive, making it easy to distinguish. However, the model performed somewhat poorer on the children playing and street music categories due to either extremely quiet sounds or loud background noises in the audio files, making it hard for even a human to distinguish.

Lastly, Table~\ref{tab:confusion} displays a confusion matrix. Any nonzero value that falls outside of the left diagonal line represents files the model classified incorrectly. In this case, RALCT classified only 96 of 8,721 audio files incorrectly, which is only about 11\%.

\section*{Conclusion}

In this paper, we proposed a novel and lightweight machine learning framework as a solution to improve the lives of the hearing impaired. Compared to hearing aids, our solution is more reliable and affordable. Compared to existing ESC architectures, RALCT utilizes less resources due to its small number of parameters and fast training time, making it suitable for mobile integration while achieving consistent state-of-the-art (SOTA) performance for over ten versions.

\subsection*{Limitations and Improvements}

We suspect that the augmentation chance variables may have limited RALCT's performance: setting the probabilities to a value higher than 0.2 led to frequent RAM overload and session crashes in Colab, preventing RALCT from being exposed to a higher variety of sound modifications that could help our model generalize between classes better. We suspect using a dedicated environment, such as Linux, that contains more RAM can potentially achieve higher performance.

\subsection*{Mobile App Integration}

RALCT is one of the first ever models developed specifically to be lightweight and accurate enough to deploy to mobile devices, with only 310K parameters and taking up only 2.8 MB of space as a TensorFlow Lite Model. The TensorFlow Lite framework is shown to be versatile for numerous mobile development software, e.g., Android Studio, Flutter, CoreML, etc. for iOS and Android.

\begin{figure}[htbp]
\centering
\includegraphics[width=\linewidth]{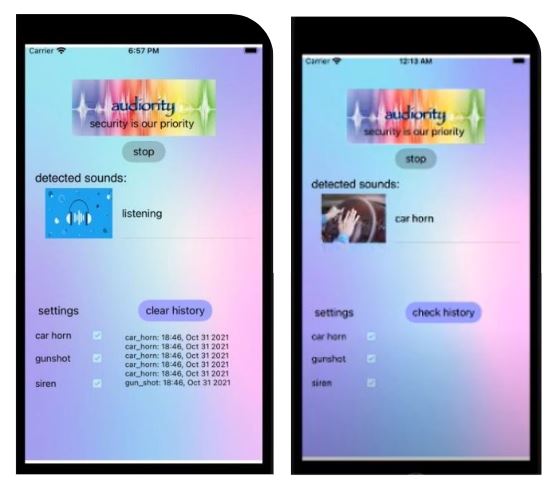}
\caption{Two screenshots of our proposed mobile app, Audiority. The left picture displays the history of the car horn sound, and the ``listening'' icon, which means the phone's microphone is actively looking for surrounding sounds but hasn't detected one yet. The right image displays the ``car horn'' category and a picture of the car horn, meaning that Audiority has detected a car horn nearby. We tested this app with YouTube sound clips that were played nearby.}
\label{fig:app}
\end{figure}

We built our iOS mobile app prototype, Audiority, using Swift, CreateML, and XCode. To test real-world sounds, we set up a phone's built-in microphone to listen for nearby sounds in real time. Currently, Audiority can detect YouTube testing clips containing the most critical sound categories (car horn, gunshot, and siren) within 1 second, which is the current inference time. When a sound is detected, the app immediately displays the name of the sound along with its pictorial representation in case the user does not speak English. Users of our app can also record the sound detection history, which includes the time, date, and name of the sound based on users' preferences for reporting purposes. Screenshots of Audiority detailing these functions are displayed in Figure~\ref{fig:app}.

Future work involves developing an Android version to reach out to more hearing-impaired users. Next, we plan to partner with hearing-impaired people to test our solution in real-world scenarios.

\subsection*{Summary and Significance}

RALCT is the first ever lightweight algorithm created solely for the purpose of improving outdoor safety for the hearing impaired and is suitable for mobile deployment and integration. We also developed a phone app that can integrate with our model to display predicted sound categories on the phone screen to alert people of potential dangers in real-time environments. We believe this research has a great potential to advance into a viable solution on the market to assist the hearing-impaired community.

\section*{Acknowledgments}

The author wishes to thank Dr. Thomas Hymel of Stanford University for his guidance in reviewing and proofreading machine learning experiments.


\begin{thebibliography}{99}

\bibitem{abdoli2019}
Abdoli, S., Cardinal, P., \& Koerich, A. (2019). ``End-to-end environmental sound classification using a 1D convolutional neural network,'' \emph{Expert Systems with Applications}, vol.~136, pp.~252--263.

\bibitem{ahmadian2021}
Ahmadian, A., Yang, L., Zhao, H. (2021). Sound Classification Based on Multihead Attention and Support Vector Machine. \emph{Hindawi}, Volume 2021. \url{https://doi.org/10.1155/2021/9937383}.

\bibitem{allard2019}
Allard, M. (2019). What is a Transformer? \emph{Medium}. \url{https://medium.com/inside-machine-learning/what-is-a-transformer-d07dd1fbec04}

\bibitem{audiologist2020}
Audiologist, C. L. (2020). Why does the world sound so noisy with hearing aids? \emph{Clear Living}. \url{https://www.clearliving.com/hearing/hearing-aids/background-noise/}.

\bibitem{boddapati2017}
Boddapati, V., Petef, A., Rasmusson, J., \& Lundberg, L. (2017). ``Classifying environmental sounds using image recognition networks,'' \emph{Procedia Computer Science}, vol.~112, pp.~2048--2056.

\bibitem{bonner2019}
Bonner, A. (2019). The complete beginner's guide to deep learning: Convolutional Neural Networks. \emph{Medium}. \url{https://towardsdatascience.com/wtf-is-image-classification-8e78a8235acb}

\bibitem{dai2021}
Dai, Y., Gao, Y., \& Liu, F. (2021). TransMed: Transformers Advance Multi Modal Medical Image Classification. \emph{Diagnostics (Basel, Switzerland)}, 11(8), 1384. \url{https://doi.org/10.3390/diagnostics11081384}

\bibitem{fourie2018}
Fourie, C. (2018). Hearing aids don't work for everyone. \emph{Hearing Aid Specialists in Australia}. \url{https://www.valuehearing.com.au/news/why-hearing-aids-dont-always-work-well}

\bibitem{giacaglia2019}
Giacaglia, G. (2019). Transformers. \emph{Medium}. \url{https://towardsdatascience.com/transformers-141e32e69591}

\bibitem{gorgolewski2020}
Gorgolewski, C. (2020). UrbanSound8K. \emph{Kaggle}. \url{https://www.kaggle.com/chrisfilo/urbansound8k}

\bibitem{gupta2017}
Gupta, D. (2017). Transfer learning: Pretrained models in Deep Learning. \emph{Analytics Vidhya}. \url{https://www.analyticsvidhya.com/blog/2017/06/transfer-learning-the-art-of-fine-tuning-a-pre-trained-model/}

\bibitem{guzhov2021}
Guzhov, A., Raue, F., Hees, J., \& Dengel, A. (2021). Audioclip: Extending clip to image, text and audio. \emph{arXiv preprint arXiv:2106.13043}.

\bibitem{kazemnejad2019}
Kazemnejad, A. (2019). Transformer architecture: The positional encoding. \emph{Transformer Architecture: The Positional Encoding - Amirhossein Kazemnejad's Blog}. \url{https://kazemnejad.com/blog/transformer_architecture_positional_encoding/}

\bibitem{li2018}
Li, S., Yao, Y., Hu, J., Liu, G., Yao, X., \& Hu., J. (2018). ``An ensemble stacked convolutional neural network model for environmental event sound recognition,'' \emph{Applied Science}, vol.~8, no.~7, pp.~1--20.

\bibitem{olson2021}
Olson, S. (2021). How much do hearing aids cost in 2022? \emph{Health.com}. \url{https://www.health.com/health-reviews/hearing-aid-prices}

\bibitem{nandi2021}
Nandi, P. (2021). CNNs for Audio Classification. \emph{Medium}. \url{https://towardsdatascience.com/cnns-for-audio-classification-6244954665ab}

\bibitem{palanisamy2020}
Palanisamy, K., Singhania, D., Yao, A. (2020). Rethinking CNN Models for Audio Classification. \emph{arXiv preprint arXiv:2007.11154v2}.

\bibitem{salamon2016}
Salamon, J., \& Bello, J.~P. (2016). Deep convolutional neural networks and data augmentation for environmental sound classification. \emph{IEEE Signal Processing Letters}, 24(3), 279--283.

\bibitem{vaswani2017}
Vaswani, A., Shazeer, N., Parmar, N., Uszkoreit, J., Jones, L., Gomez, A.~N., \& Polosukhin, I. (2017). Attention is all you need. In \emph{Advances in Neural Information Processing Systems} (pp.~5998--6008).

\bibitem{who}
World Health Organization. (n.d.). Deafness and hearing loss. \emph{World Health Organization}. \url{https://www.who.int/health-topics/hearing-loss}

\bibitem{you2019}
You, K., Long, M., Wang, J., \& Jordan, M.~I. (2019). How does learning rate decay help modern neural networks? \emph{arXiv preprint arXiv:1908.01878}.

\bibitem{zhang2018}
Zhang, Z., Xu, S., Cao, S., \& Zhang, S. (2018). Deep convolutional neural network with mixup for environmental sound classification. In \emph{Chinese Conference on Pattern Recognition and Computer Vision (PRCV)} (pp.~356--367).

\end{thebibliography}
\end{document}